\documentclass[doublespacing]{elsart}

\usepackage{graphicx}
\usepackage{amssymb}

\begin{document}
\begin{frontmatter}


\journal{SCES'2001: Version 1}


\title{Photoemission  view of    electron fractionalization in   quasi-one
dimensional metal Li$_{0.9}$Mo$_6$O$_{17}$}

%
%
%
%
%
%

\author[MI]{G.-H. Gweon\corauthref{1}}
\author[ALS]{J.D. Denlinger}
\author[AMES]{C.G. Olson}
\author[SRC]{H. H\"{o}chst}
\author[CNRS]{J. Marcus}
\author[CNRS]{C. Schlenker}

%
 
\address[MI]{Randall Lab.\ of Physics, Univ.\ of Michigan, Ann
  Arbor, MI 48109, USA}
\address[ALS]{Adv.\ Light Source, Lawrence Berkeley National Lab.,
  Berkeley, CA 94720, USA}
\address[AMES]{Ames Laboratory, Iowa State University, Ames, Iowa 50011,
  USA}
\address[SRC]{Synchrotron Radiation Center, Univ.\ of Wisconsin,
  Stoughton, WI 53589, USA}
\address[CNRS]{LEPES -- CNRS, BP166, 38042 Grenoble Cedex, France}

%
%
%
%


%
%
%
%

\corauth[1]{Corresponding Author: Randall Lab.\ of Physics, Univ.\ of
  Michigan, Ann Arbor, MI 48109-1120, USA. Phone: (734) 647-9434, Fax:
  (734) 763-9694, Email: gweon@umich.edu}


\begin{abstract}
 
  We report Luttinger liquid (LL) line shapes better revealed by new angle
  resolved photoemission data taken  with a much improved angle resolution
  on  a quasi-1-dimensional metal  Li$_{0.9}$Mo$_6$O$_{17}$.  The new data
  indicate a   larger spinon velocity  than our  previous lower resolution
  data indicated.

\end{abstract}

%
%



\end{frontmatter}

%
%
%
%
%

Of  central interest  in    current   condensed  matter physics    is  the
fractionalization \cite{orgad}   of an electron  into density   waves in a
quasi one dimensional  (quasi-1D) metal, e.g.  spinons  and holons  in the
Luttinger liquid  (LL), and dimensional  crossover \cite{carlson} in which
an  electron  (or a  Landau  quasi-electron)  re-emerges as  the effective
dimension is changed  to  three, e.g.   by  lowering the  temperature  (T)
through a  phase transition.    Angle resolved photoemission  spectroscopy
(ARPES) is an ideal tool to study these phenomena through  the view of the
single  particle       spectral    function,   and     we     have   shown
\cite{denlinger,gweon-comment,gweon-jesrp}  that  the     quasi-1D  metals
Li$_{0.9}$Mo$_6$O$_{17}$  and K$_{0.3}$MoO$_{3}$ are  very  good materials
for such studies.  In particluar, the  former is suitable for studying the
fractionalization phenomenon  and  the  latter  for  the   the dimensional
crossover  phenomenon.  In this paper, due  to the space limit, we discuss
Li$_{0.9}$Mo$_6$O$_{17}$ only.

In our  previous  work  \cite{denlinger}  on  Li$_{0.9}$Mo$_6$O$_{17}$, we
reported non-Fermi  liquid   (non-FL)  ARPES line shapes   which  compared
favorably with an  LL theory.  In  this paper, we  analyze data taken with
better angular resolution (0.36$^o$ instead of 2$^o$), which give a better
view of LL line shapes.  Therefore, this paper  can be viewed as providing
details  of our  previous work \cite{gweon-comment,gweon-jesrp} disproving
another  group's claim \cite{smith} that  FL  line shapes are revealed  by
improved angle  resolution  in Li$_{0.9}$Mo$_6$O$_{17}$.   Fig.\  1  shows
ARPES data taken along the $\Gamma$-Y line at T = 250 K with 30 eV photons
at the 4m NIM beam line equipped  with an SES 200  Scienta analyzer at the
Synchrotron  Radiation Center (SRC).   The   energy resolution was 49  meV
FWHM, practically  the  same  as   before \cite{denlinger}.  This   energy
resolution is somewhat worse than what was  used for other high resolution
data   published  already  \cite{gweon-comment,gweon-jesrp}, but   gives a
better  signal to noise  ratio without compromising the  line shapes.  The
bands A,B,C,D found   in  Fig.\ 1 correspond  very   well to the   4 bands
predicted by band theory \cite{whangbo}, consistent with our more detailed
band structure analysis \cite{gweon-jesrp}.  The band D, observed strongly
along a different $k$-path \cite{denlinger}, is very weak  in Fig.\ 1, and
the band C is observed more clearly due to the improved angular resolution
\cite{gweon-comment}.   As a result, we can  now study $E_F$ crossing line
shapes of a single 1D band.

We  focus our attention on   the line shapes  of  band  C  near its  $E_F$
crossing in Fig.\  2   (a).   The band B,    while  strong in   intensity,
contributes little weight  at energies close  to or greater  than the peak
positions  of the band   C, according to our test   modelling of  its line
shape.  Therefore, the band  B distorts only the  high binding energy tail
region  of the band C  line shapes.  The current data  are  similar to our
lower resolution  data \cite{denlinger} in showing the  peak and  the edge
moving at different  speeds, a monotonic decrease  of the peak height, and
intensity   retraction  from $E_F$  for  $k   >  k_F$.  However, there are
important differences.  (1) $E_F$ weight  relative to the peak height  for
$k = k_F$ is now much larger.  (2) As the inset shows, the edge is steeper
and  in consequence the movement  of  the zero intensity  intercept of the
edge, taken as  indicative of  the spinon  velocity,  is now  only  twice,
instead of five times, slower than the peak.

As we argued before  \cite{gweon-comment}, (1) is  a simple consequence of
the  improved  angular  resolution.   We   note that  the  increased $E_F$
intensity is still  much less than that  expected for a FL, marginal Fermi
liquid  (MFL)  \cite{varma}  or  a   small   anomalous dimension  $\alpha$
($\approx  0$)  LL, as  demonstrated by  an MFL   simulation with coupling
constant  = 1 and $\omega_c  = 1$ eV, shown in  the inset of  Fig.\ 2 (b).
Another reason to  reject these ground states is  that the peak sharpening
expected in  their line shapes as $k  \rightarrow k_F$  is not observed in
our data.  In fact, the line  shape at $k  = k_F$ is very featureless. The
$k = k_F$ line shape is well  explained within the LL,  as seen in Fig.\ 2
(b) which  shows an LL  simulation within  a Tomonaga-Luttinger (TL) model
\cite{meden} with $\alpha = 0.9$ for which $\beta$, the ratio of the holon
velocity to  the spinon velocity,  is 5.   This is  the same simulation as
that presented by  us   previously \cite{denlinger} except for  using  the
improved angle resolution of the new  data and the unimportant differences
of  T (for Fermi-Dirac cutoff) and  $v_F$ (0.8 eV   \AA, instead of 0.7 eV
\AA, describes the peak movement better in this narrow momentum range).

The finding (2)  is  not explained  by  our simulation, because the   kink
between the holon peak and the spinon edge and the slow spinon edge in the
simulation are  not observed in  the data.  Nonetheless, the more gap-like
line shape  produced by  the better resolved  steeper  edge is   even more
suggestive of the generic LL line  shape with its  gap around $E_F$ for $k
\ne k_F$.  Finding (2) implies a $\beta$ value more like  2 instead of the
value 5 inferred previously, for which $\alpha  = 1/8$ would obtain in the
TL model used.  However, $\alpha = 1/8$ would imply  a spinon peak instead
of an  edge,  and  the ($k_F$,   $E_F$)  weight is  strongly  dependent on
$\alpha$, and so we must take the $\alpha$ value to be unchanged. Thus the
better resolved spectra show  more clearly the essential  Non-Fermi liquid
character but force us to abandon the $\alpha$-$\beta$ relationship of the
model.  Because this relationship is known to  be highly special to the TL
model   used and subject  to  modification  due  to  residual interactions
present  in  other models, the   1D  Hubbard model  \cite{schulz}  being a
pointed example, it is easily given up.

In conclusion,  we  have  reported LL   line shapes  revealed by  new high
resolution        ARPES    data  on Li$_{0.9}$Mo$_6$O$_{17}$.   Currently,
Li$_{0.9}$Mo$_6$O$_{17}$   remains as the best   example  to show LL  line
shapes.  We  note that the theory  of  LL line shapes  is making  a steady
progress.   For example, a  finite  T  theory \cite{orgad-finiteT} is  now
available. An identification  of finite T  features in  our data  set is a
work in progress.

The  work  at U-M   was   supported by the   U.S.   NSF  under  the  grant
DMR-99-71611) and the U.S.  DoE by the grant DE-FG-02-90ER45416.  The Ames
Lab is  supported by the US DOE  under Contract No.\ W-7405-ENG-82 and the
SRC is supported by the US NSF Grant No.\ DMR-00-84402.

%
%
%
%

%
%

\begin{figure}
    \centering
    \includegraphics{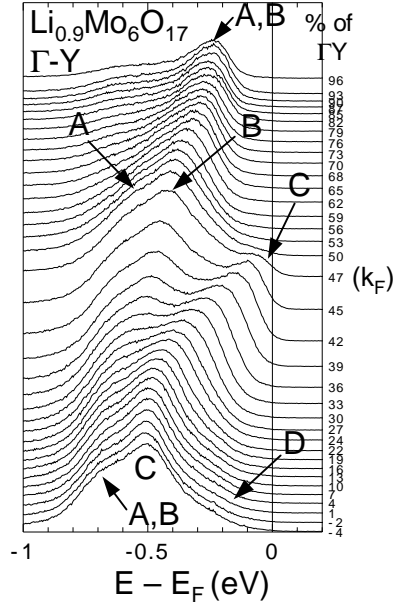}
    \caption{ARPES data on  Li$_{0.9}$Mo$_6$O$_{17}$. See text for details.}
\end{figure} 

\begin{figure}
    \centering \includegraphics{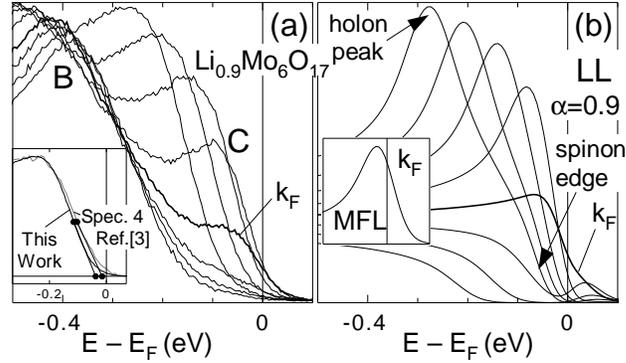}
    \caption{(a) ARPES data on  Li$_{0.9}$Mo$_6$O$_{17}$ near $k = k_F$
      and (b) LL line shape simulation. See text for details.}
\end{figure}  
%
%


\end{document}